\documentclass[floats,floatfix,showpacs,amssymb,prd,twocolumn,superscriptaddress,nofootinbib]{revtex4-1}

\usepackage{graphicx,epsf, epsfig, amssymb}
\usepackage{bm}
\usepackage{longtable}
\usepackage{color}
\usepackage[breaklinks]{hyperref}
\usepackage{amsfonts,amsmath,amssymb,mathrsfs}

\def\be{\begin{equation}}
\def\ee{\end{equation}}
\def\beq{\begin{eqnarray}}
\def\eeq{\end{eqnarray}}
\def\f{\frac}
\def\p{\partial}
\newcommand{\nn}{\nonumber}
\def\ii{{\rm i}}

\newcommand{\bgaln}{\begin{align}}
\newcommand{\bgeq}{\begin{equation}}

\newcommand{\Rmnum}[1]{\expandafter\@slowromancap\romannumeral #1@}

\begin{document}

\title {Light scalar field constraints from gravitational-wave
  observations of compact binaries}

\author{Emanuele Berti} \email{berti@phy.olemiss.edu}
\affiliation{Department of Physics and Astronomy, The University of
  Mississippi, University, MS 38677, USA}
\affiliation{California Institute of Technology, Pasadena, CA 91109, USA}

\author{Leonardo Gualtieri} \email{Leonardo.Gualtieri@roma1.infn.it}
\affiliation{Dipartimento di Fisica, Universit\`a di Roma ``Sapienza''
  \& Sezione, INFN Roma1, P.A. Moro 5, 00185, Roma, Italy.}

\author{Michael Horbatsch} \email{horbatm@univmail.cis.mcmaster.ca}
\affiliation{Department of Physics and Astronomy, McMaster University,
  1280 Main St. W, Hamilton, Ontario, Canada, L8S 4L8.}

\author{Justin Alsing} \email{justin.alsing@seh.ox.ac.uk}
\affiliation{Department of Physics, University of Oxford, Keble Road,
  Oxford OX1 3RH, UK}

\date{\today}

\begin{abstract}
Scalar-tensor theories are among the simplest extensions of general
relativity. In theories with light scalars, deviations from Einstein's
theory of gravity are determined by the scalar mass $m_s$ and by a
Brans-Dicke-like coupling parameter $\omega_{\rm BD}$. We show that
gravitational-wave observations of nonspinning neutron star-black hole
binary inspirals can be used to set lower bounds on $\omega_{\rm BD}$
and upper bounds on the combination $m_s/\sqrt{\omega_{\rm BD}}$. We
estimate via a Fisher matrix analysis that individual observations
with signal-to-noise ratio $\rho$ would yield $(m_s/\sqrt{\omega_{\rm
    BD}})(\rho/10)\lesssim 10^{-15}$, $10^{-16}$ and $10^{-19}$~eV for
Advanced LIGO, ET and eLISA, respectively. A statistical combination
of multiple observations may further improve these bounds.
\end{abstract}
\maketitle

\section{Introduction}
\label{sec:intro}

Scalar-tensor theories, in which gravity is mediated by a tensor field
as well as a nonminimally coupled scalar field, are popular and simple
alternatives to Einstein's general relativity
\cite{Fujii:2003pa,EspositoFarese:2009ta,Clifton:2011jh}. Generic
scalar-tensor theories are of interest in cosmology, and under
certain conditions they can be shown to be equivalent to $f(R)$
theories \cite{Sotiriou:2008rp,DeFelice:2010aj}; they have also been
investigated in connection with inflation and cosmological
acceleration (see e.g. \cite{Steinhardt:1994vs,Boisseau:2000pr}).

Light, massive scalars are a generic prediction of the low-energy
limit of many theoretical attempts to unify gravity with the Standard
Model of particle physics. For example, string theory suggests the
existence of light scalars (``axions'') with masses $m_s$ that could
be as small as the Hubble scale ($\sim 10^{-33}$~eV).  For a coherent
scalar field to be cosmologically distinct from cold dark matter, or
to play a quintessence-like role, it must be very light:
$10^{-33}~{\rm eV}<m_s<10^{-18}~{\rm eV}$.  In this ``string
axiverse'' scenario, cosmic microwave background observations, galaxy
surveys and even astrophysical measurements of black hole spins may
offer exciting experimental opportunities to set constraints on the
masses and couplings of the scalar fields
\cite{Arvanitaki:2010sy,Kodama:2011zc,Marsh:2011bf}. Therefore,
astrophysical and cosmological observations may provide important
clues on the relation between gravity and the other forces.

In this paper, building on previous work \cite{Alsing:2011er}, we
explore the possibility of constraining the mass and coupling of a
scalar field via gravitational-wave observations of compact binary
inspirals with Earth-based or space-based interferometers, such as
Advanced LIGO/Virgo \cite{Shoemaker:aLIGO,2010CQGra..27h4006H,aVIRGO},
the Einstein Telescope \cite{ET,2010CQGra..27s4002P}, eLISA/NGO
\cite{AmaroSeoane:2012km} or Classic LISA \cite{Danzmann:1998}.
Until recently, calculations of gravitational radiation in
scalar-tensor theories focused on massless scalar fields (see
e.g.~\cite{Wagoner:1970vr,Will:1989sk,Shibata:1994qd,Harada:1996wt,Brunetti:1998cc,Damour:1998jk,Sotani:2005qx}). When
$m_s=0$, deviations from general relativity can be parametrized in
terms of a single coupling constant, that for historical reasons is
usually chosen to be the Brans-Dicke parameter $\omega_{\rm BD}$. The
best constraint on this parameter to date ($\omega_{\rm
  BD}>\omega_{\rm Cass}=40,000$) comes from Cassini measurements of
the Shapiro time delay \cite{Will:2005va}. For massless scalar-tensor
theories, gravitational-wave observations of mixed binaries can
provide constraints comparable to (or marginally better than) the
Cassini bound \cite{Will:1994fb,Berti:2004bd,Yagi:2009zm}.
Recent studies explored some of the implications that light, massive
scalar fields may have in the context of gravitational-wave
phenomenology. For example, resonant superradiant effects induced by
light massive scalars may produce ``floating orbits'' when small
compact objects spiral into rotating black holes, leaving a distinct
signature in gravitational waves \cite{Cardoso:2011xi}.

The general action of scalar-tensor theories with a single scalar
field can be cast in the form
\begin{eqnarray}
S&=&\frac{1}{16\pi}\int\left[\phi R-\frac{\omega(\phi)}{\phi}g^{\mu\nu}
\phi_{,\mu}\phi_{,\nu}+M(\phi)\right](-g)^{\frac{1}{2}}d^4x\nonumber\\
&&+\int{\cal L}_M(g^{\mu\nu},\Psi)\label{action}\,,
\end{eqnarray}
where ${\cal L}_M$ is the matter Lagrangian and $\Psi$ collectively
denotes the matter fields. In this paper we consider a scalar-tensor
theory with a constant coupling function $\omega(\phi)=\omega_{\rm
  BD}$ and a mass term, as in \cite{Alsing:2011er}. Expanding the
potential $M(\phi)$ about a (cosmologically determined) background
value of the scalar field $\phi=\phi_0$, under the assumption of
asymptotic flatness, we have
\begin{equation}
M(\phi)\simeq\frac{1}{2}M''(\phi_0)(\phi-\phi_0)^2\,.
\end{equation}
Then the equation of motion for the scalar field has the schematic
form \cite{Alsing:2011er}
\begin{equation}
\Box_g\phi-m_s^2(\phi-\phi_0)=\hbox{(source terms)}\,,
\end{equation}
where
\begin{equation}
m_s^2=-\frac{\phi_0}{3+2\omega_{\rm BD}}M''(\phi_0)\,.
\end{equation}
We note that for $m_s=0$ this theory reduces to the well-known
Brans-Dicke theory \cite{Brans:1961sx}.

Throughout the paper we use geometrical units $G=c=1$, so all
quantities can be expressed in (say) seconds: for example, the mass of
the sun $M_\odot=4.926\times 10^{-6}$~s. We assume a standard
$\Lambda$CDM cosmological model with $H_0=72$~km\,s$^{-1}$Mpc$^{-1}$,
$\Omega_M=0.3$ and $\Omega_\Lambda=0.7$. Given the scalar field mass
$\tilde m_s$ (in eV), we find it convenient to define a quantity
$m_s=\tilde m_s/\hbar$ with dimensions of inverse length (or inverse
time, since $c=1$), as this is the quantity that would appear in the
flat-space Klein-Gordon equation $(\Box-m_s^2)\varphi=0$. To convert
between $\tilde m_s$ and $m_s$, it is sufficient to note that
$\hbar=6.582\times 10^{-16}$~eV\,s. We will use $\tilde m_s$ and $m_s$
interchangeably in the rest of the paper; the units should be clear
from the context.

Our study is the first to explore bounds on massive scalar tensor
theories using gravitational-wave observations of compact binaries.
We only consider quasicircular, nonspinning binaries for two reasons:
(1) resonant effects in the context of gravitational-wave detection
were considered in \cite{Yunes:2011aa}, and they would produce large
enough dephasing to prevent detection using general relativistic
templates; (2) in the absence of resonant effects, the introduction of
aligned spins would increase parameter measurement errors by about one
order of magnitude. However, spin precession and eccentricity would
reduce the errors by a comparable amount
\cite{Stavridis:2009mb,Yagi:2009zm}. For these reasons, our bounds for
quasicircular, nonspinning binaries should be comparable to bounds
resulting from more realistic (and complex) waveform models, as long
as resonant effects don't play a role.

In this paper we will focus on quasicircular neutron star-black hole
binaries.  The main reason is that, according to our previous
investigation \cite{Alsing:2011er}, only {\it mixed binaries} (i.e.,
binaries whose members have different gravitational binding energies)
can produce significant amounts of scalar gravitational
radiation. Under the assumption of asymptotic flatness, dipole
radiation is produced due to violations of the strong equivalence
principle when the binary members have unequal ``sensitivities'':
$s_1\neq s_2$.  The sensitivities are related to the gravitational
binding energies of each binary member (labeled by $a=1\,,2$):
$s_a=1/2$ for a Schwarzschild black hole, $s_a\sim0.2$ for a neutron
star, and $s_a\sim10^{-4}$ for a white dwarf
\cite{Will:1989sk,Zaglauer:1992bp}. Roughly speaking, dipole radiation
is produced when the system's center of mass is offset with respect to
the center of inertia, so mixed and eccentric binaries are the best
observational targets to constrain scalar-tensor theories. A second
reason to consider neutron star-black hole binaries is that dipolar
radiation should not be emitted in black hole-black hole systems
because of the no-hair theorem, i.e. the fact that black hole
solutions in scalar-tensor theories are the same as in GR (see
\cite{Sotiriou:2011dz} and references therein). Recently, building on
earlier work by Jacobson \cite{Jacobson:1999vr}, Horbatsch and Burgess
pointed out that scalar fields that vary on cosmological timescales
may violate the no-hair theorem, so that even black hole-black hole
binaries may produce dipolar radiation \cite{Horbatsch:2011ye}.  The
existence of dipolar radiation in extended theories of gravity and the
investigation of possible bounds on dipolar radiation are active
research topics \cite{Yagi:2011xp,Arun:2012hf,Chatziioannou:2012rf}.

In summary, here we will make the conservative assumption that only
mixed binaries generate dipolar radiation, and we will investigate
bounds on the mass and coupling of the scalar field coming from
gravitational-wave observations of black hole-neutron star
binaries. For simplicity we will set the mass of the neutron star to
be $M_{\rm NS}=1.4~M_\odot$, and we will focus on nonrotating black
holes. This rules out by construction the possibility of floating
orbits of the kind studied in \cite{Cardoso:2011xi,Yunes:2011aa}.

We will consider the bounds that could be obtained using different
gravitational-wave detectors. For Advanced LIGO we use the fit to the
expected power spectral density given in Table I of
\cite{Sathyaprakash:2009xs} (henceforth ``AdLIGO'') as well as the
“zero-detuning, high power” configuration, as given analytically in
Eq.~(4.7) of \cite{Ajith:2011ec} (``AdLIGO ZDHP''). In both cases we
assume the power spectral density to be infinite below a seismic noise
cutoff frequency of 20~Hz. For the Einstein Telescope we use the
analytical fit presented in Table I of \cite{Sathyaprakash:2009xs},
assuming a lower cutoff frequency of 10~Hz.  For Classic LISA we use
the analytical Barack--Cutler expression \cite{Barack:2003fp}, as
corrected in \cite{Berti:2004bd}.  For eLISA/NGO we use the noise
model (inclusive of galactic background noise) discussed in
\cite{AmaroSeoane:2012km}.

In general, the gravitational radiation from a binary in massive
scalar-tensor theories depends on the scalar field mass $m_s$ and on
the coupling constant $\xi\sim 1/\omega_{\rm BD}$ (see
\cite{Alsing:2011er} and Eq.~(\ref{definitions}) below for a more
precise definition).  
Matched-filtering detection of gravitational radiation from compact
binaries relies primarily on an accurate knowledge of the wave
phasing. In any theory that differs from general relativity, the
gravitational-wave phasing will be modified by a slight amount that
depends on the parameters of the theory. In massless scalar-tensor
theories the dominant (dipolar) contribution scales like
$1/\omega_{\rm BD}$. If we measure a waveform consistent with general
relativity we can set upper bounds on deviations of the dipolar term
from zero, and therefore we can set lower limits on $\omega_{\rm BD}$
\cite{Will:1994fb,Berti:2004bd,Yagi:2009zm}.
By computing the gravitational-wave phase in the stationary-phase
approximation (SPA; cf. Eq.~(\ref{spafeta2}) below) we will see that
the scalar mass always contributes to the phase in the combination
$m_s^2 \xi\sim m_s^2/\omega_{\rm BD}$, so that gravitational-wave
observations of nonspinning, quasicircular inspirals can only set
upper limits on $m_s/\sqrt{\omega_{\rm BD}}$. For large
signal-to-noise ratio (SNR) $\rho$, the constraint is inversely
proportional to $\rho$. It is useful to plot constraints that result
from observations with SNR $\rho=10$: this corresponds to a minimal
threshold for detection of the corresponding binary system, and in
this sense it places an upper limit on the achievable bounds on $m_s$
(and a lower limit on the achievable bounds on $\omega_{\rm BD}$).

\begin{figure}[thb]
\includegraphics[width=8.5cm,clip=true]{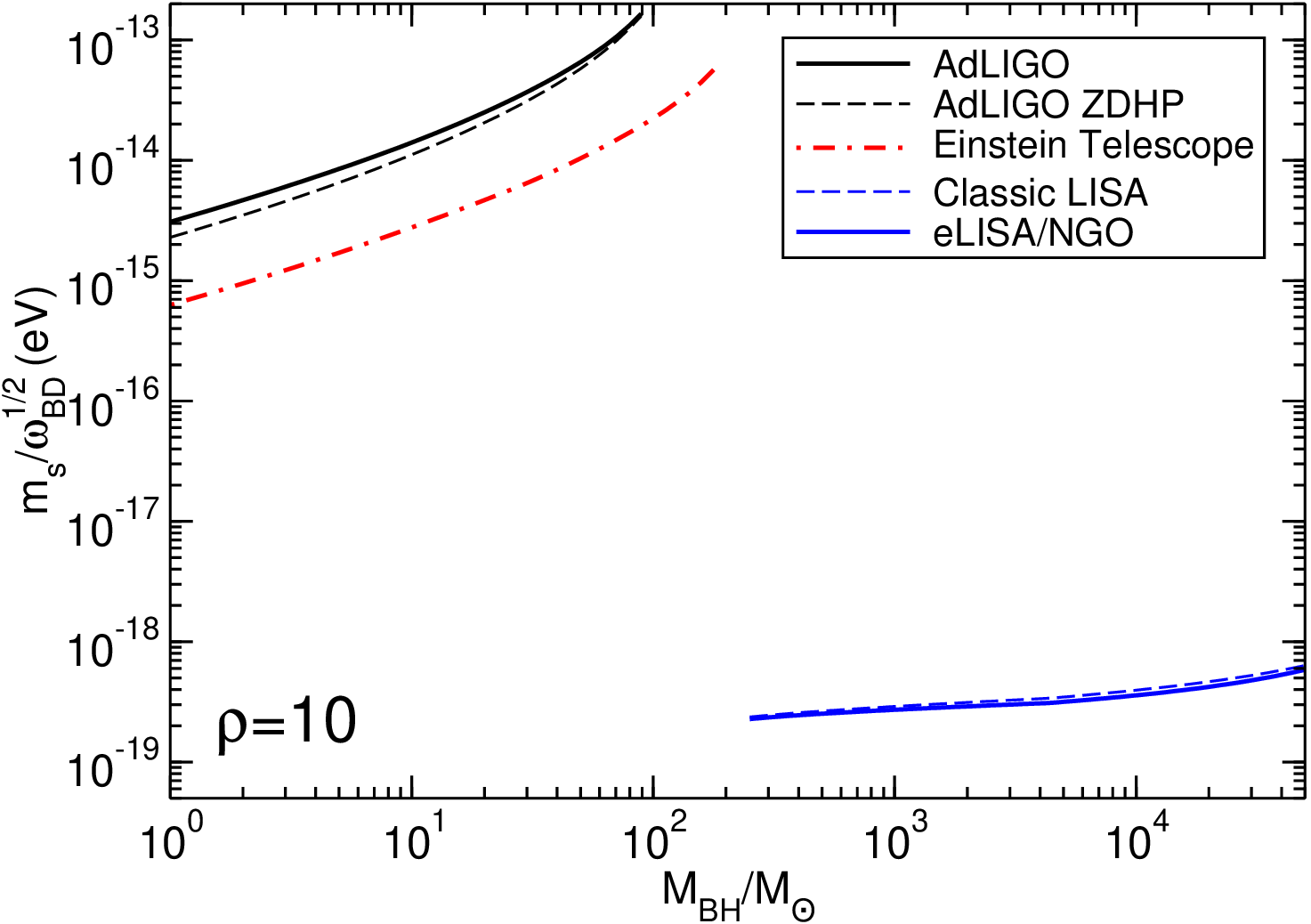}
\caption{Bounds on $m_s/\sqrt{\omega_{\rm BD}}$ with AdLIGO, AdLIGO
  ZDHP, ET, Classic LISA and eLISA/NGO at fixed SNR $\rho=10$.}
\label{fig:mswBD}
\end{figure}

In Fig.~\ref{fig:mswBD} we plot the upper bound on
$m_s/\sqrt{\omega_{\rm BD}}$ resulting from neutron star-black hole
binary observations with SNR $\rho=10$, as a function of the black
hole mass $M_{\rm BH}$, for different detectors.  The order of
magnitude of these bounds is essentially set by the lowest frequency
accessible to each detector, and it can be understood by noting that
the scalar mass and gravitational-wave frequency are related (on
dimensional grounds) by $m_s({\rm eV})=6.6\times 10^{-16}\, f({\rm
  Hz})$, or equivalently $f({\rm Hz})=1.5\times 10^{15}\, m_s({\rm
  eV})$.
%
For eLISA, the lower cutoff frequency (imposed by acceleration noise)
$f_{\rm cut}\sim 10^{-5}$~Hz corresponds to a scalar of mass
$m_s\simeq 6.6\times 10^{-21}$~eV. For Earth-based detectors the
typical seismic cutoff frequency is $\sim 10$~Hz, corresponding to
$f_{\rm cut}\sim 6.6\times 10^{-16}$~eV. These lower cutoff
frequencies set the order of magnitude of scalar masses probed by 
space-based and Earth-based detectors. 

The best bounds are obtained from the intermediate mass-ratio inspiral
of a neutron star into a black hole of mass $M_{\rm BH}\lesssim
10^3~M_\odot$, as observed by a space-based instrument such as eLISA
or Classic LISA. In summary, we conclude that the most competitive
bounds would come from space-based gravitational wave detectors, and
that they would be of the order
\be 
\left(\f{m_s}{\sqrt{\omega_{\rm BD}}}\right)
\left(\f{\rho}{10}\right)\lesssim 10^{-19}~{\rm eV}.
\ee
\begin{figure}[thb]
\includegraphics[width=8.5cm,clip=true]{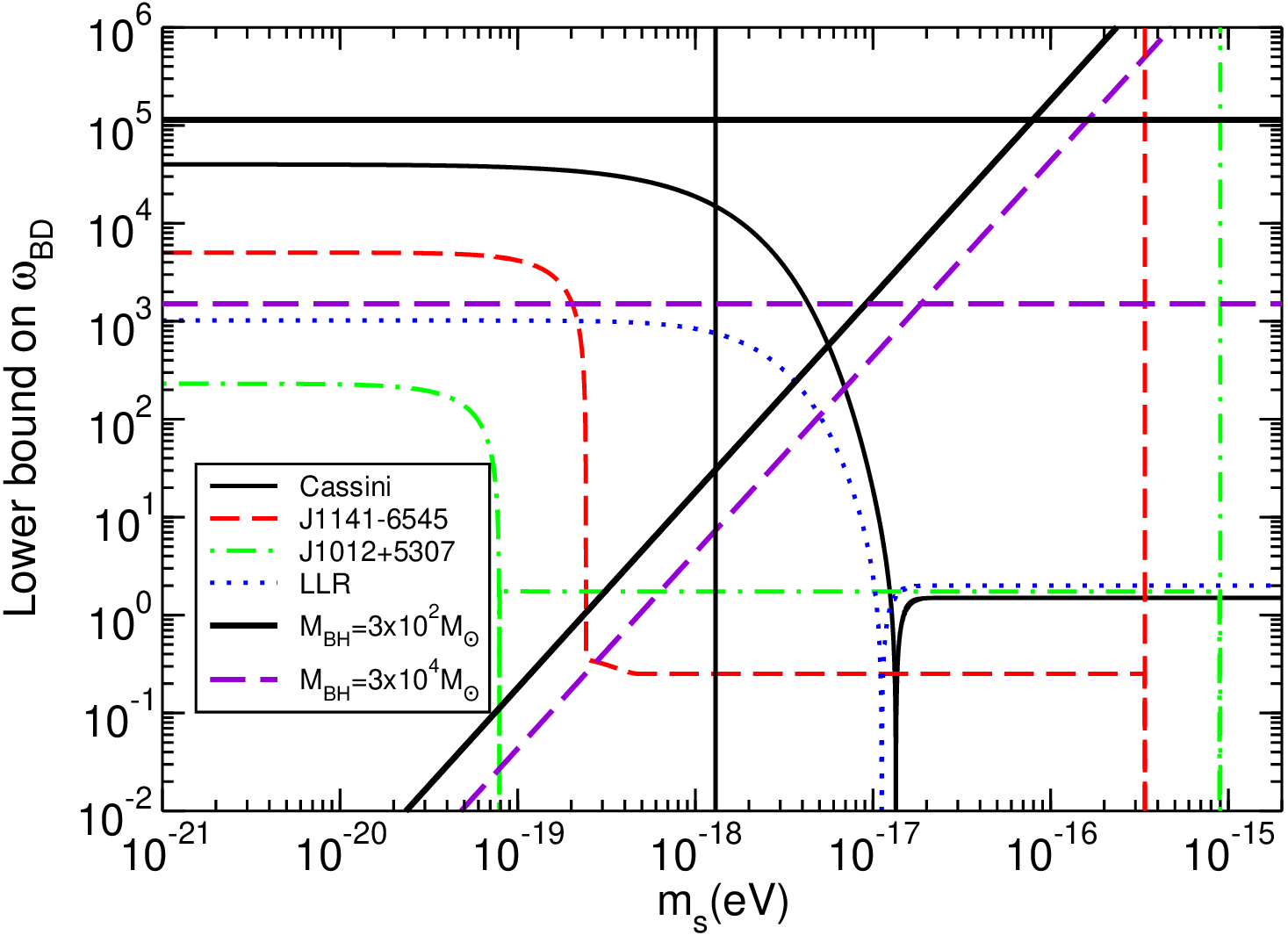}
\caption{Bounds from eLISA/NGO at SNR $\rho=10$, compared to Solar
  System and binary pulsar bounds.}
\label{fig:omega}
\end{figure}

In Fig.~\ref{fig:omega} we plot current bounds on $\omega_{\rm BD}$ as
a function of $m_s$ coming from (i) Cassini measurements of the
Shapiro time delay (solid black line), (ii) Lunar Laser Ranging bounds
on the Nordtvedt effect (dotted blue line), and (iii) measurements of
the orbital period derivative of two binary pulsar systems. We refer
the reader to \cite{Alsing:2011er} for a detailed derivation of these
bounds.

On top of the existing bounds we plot, for comparison, future bounds
from eLISA observations of neutron star-black hole binaries where the
black hole has mass $M_{\rm BH}=300~M_\odot$ (solid lines) and $M_{\rm
  BH}=3\times 10^4~M_\odot$ (dashed lines).  In both cases the bounds
refer to observations with SNR $\rho=10$.

For each system, as discussed above, gravitational wave observations
provide {\em two} constraints: a lower limit on $\omega_{\rm BD}$
(corresponding to horizontal lines in the plot) and an upper limit on
$m_s/\sqrt{\omega_{\rm BD}}$ (corresponding to the straight diagonal
lines). For each of the two neutron star-black hole systems, a
gravitational-wave observation would exclude the complement of a
trapezoidal region on the top left of the plot.  We see that bounds
from space-based gravitational-wave observations will improve over the
Cassini, Lunar Laser Ranging and binary pulsar bounds bound only for
neutron star-black hole systems with $M_{\rm BH}\lesssim
10^3~M_\odot$.  In a sense, these predictions are conservative: for
example, for a gravitational-wave observation with SNR $\rho=100$ both
bounds (the horizontal and diagonal lines) would improve by an order
of magnitude. Therefore, a single high-SNR observation (or the
statistical combination of several observations, see
e.g. \cite{Berti:2011jz}) may yield bounds on massive scalar-tensor
theories that are significantly better than Solar System or binary
pulsar bounds.

The plan of the paper is as follows. In Section~\ref{sec:SPA} we
compute the gravitational-wave phase for massive scalar-tensor
theories in the stationary phase approximation. In
Section~\ref{sec:Fisher} we use these results to compute the Fisher
information matrix in these theories. Section~\ref{sec:bounds}
presents the bounds obtainable with Earth- and space-based
interferometers. Section~\ref{sec:conclusions} discusses possible
directions for future research.

\section{Stationary phase approximation in massive scalar-tensor theories}
\label{sec:SPA}

As discussed in Section~\ref{sec:intro}, we focus on scalar-tensor
theories with constant coupling and a massive scalar field, following
the treatment and notation of \cite{Alsing:2011er}. We consider the
inspiral of a binary system composed of two compact objects with
masses $m_1$ and $m_2$. For consistency with the notation of
\cite{Berti:2004bd}, here and below all masses $m_i$ are measured in
the detector frame; they are related to the masses $m_i^{(0)}$ in the
source frame by $m_i=(1+z)m_i^{(0)}$.  Let $m=m_1+m_2$ be the total
mass of the binary, $\mu=\eta m=(m_1m_2)/(m_1+m_2)$ the reduced mass
(with $\eta$ the symmetric mass ratio), and ${\cal
  M}=\mu^{3/5}m^{2/5}=\eta^{3/5}m$ the so-called chirp
mass\footnote{Notice that Refs.~\cite{Damour:1998jk,Will:1994fb}
  introduce a slightly different definition of the chirp mass in the
  context of scalar-tensor theories.}.  We will denote by $f$ the
gravitational-wave frequency of the binary.

The dominant contributions to gravitational radiation in scalar-tensor
theories are dipolar and quadrupolar. In the adiabatic approximation,
the time derivative of the gravitational-wave frequency (denoted by an
overdot) reads \cite{Alsing:2011er}
\begin{equation}
\frac{\dot f}{f}=-\frac{\dot P}{P}=\frac{8}{5}\frac{\mu m^2}{r^4}\kappa_1
+\frac{\mu m}{r^3}{\cal S}^2\kappa_D\,,
\end{equation} 
where
\begin{align}
\xi&=\frac{1}{2+\omega_{\rm BD}}\,,\nn\\
{\cal G}&=1-\xi(s_1+s_2-2s_1s_2)\,,\nn\\
\Gamma&=1-2\frac{s_1m_2+s_2m_1}{m}\,,\nn\\
{\cal S}&=s_2-s_1\,,\nn\\
\kappa_1&={\cal G}^2
\left[12-6\xi+\xi\Gamma^2
\left(1-\frac{m_s^2}{4\pi^2f^2}\right)^2\Theta(2\pi f-m_s)\right]\,,\nn\\
\kappa_D&={\cal G}
\left[2\xi\left(1-\frac{m_s^2}{\pi^2f^2}\right)\right]
\Theta(\pi f-m_s)\,,
\label{definitions}
\end{align}
and $\Theta(x)$ is the Heaviside function.

The gravitational-wave and orbital frequencies are related by
$f=2f_{\rm orb}$, where (see e.g. Eq.~(2.25c) in \cite{Will:1989sk})
\begin{equation}
\Omega_{\rm orb}=2\pi f_{\rm orb}=\sqrt{\frac{{\cal G}m}{r^3}}\,.
\label{kepler}
\end{equation}

Following \cite{Cutler:1994ys,Poisson:1995ef}, we can compute the
phase $\psi(f)$ of the gravitational waveform as a function of the
wave frequency $f$ in the stationary phase approximation (SPA):
\begin{equation}
\psi(f)=2\pi ft(f)-\phi(f)-\frac{\pi}{4}\,,
\label{spa}
\end{equation}
where
\begin{align}
t(f)&=\int^f\frac{df'}{{\dot f}'}\,,\label{tspa}\\
\phi(f)&=2\pi\int^f\frac{f'}{{\dot f}'}df'\,.\label{phispa}
\end{align}

For light scalars ($m_s\lesssim 10^{-19}$~eV) we know that
$\omega_{\rm BD}$ must be larger than the Cassini bound $\omega_{\rm
  Cass}=40,000$ (see Fig.~\ref{fig:omega}), therefore we are justified
in assuming that $\xi\ll1$ and we can linearize in $\xi$. For scalar
masses $m_s\gtrsim 10^{-18}$~eV, strong couplings ($\xi\sim 1$ or
larger) are not experimentally ruled out; however our focus here is on
constraining light scalars of cosmological relevance. A
straightforward calculation linearizing in $\xi$ yields
\begin{align}
\dot f&=\frac{96}{5}\pi^{8/3}{\cal M}^{5/3}f^{11/3}\left\{1-\frac{2}{3}
\xi(s_1+s_2-2s_1s_2)-\frac{1}{2}\xi\right.\nn\\
&+\frac{\xi\Gamma^2}{12}\left(1-\frac{m_s^2}{4\pi^2f^2}\right)^2
\Theta(2\pi f-m_s)\nn\\
&+\frac{5}{48}\xi\left(1-\frac{m_s^2}{\pi^2f^2}\right)
{\cal S}^2(\pi mf)^{-2/3}\Theta(\pi f-m_s)+\dots\Bigg\}\,,\nn
\end{align} 
where dots denote contributions of higher post-Newtonian order.
Let us define the quantity
\begin{equation}
\nu\equiv m_s^2m^2\,.
\end{equation}
This quantity is dimensionless, since (as discussed above) the scalar
mass $m_s$ has dimensions of inverse length (or time), and in
geometrical units the total mass of the binary system has dimensions
of length (or time).

Let $m_{20}=10^{-20}$~eV be the typical scalar mass below which we get
bounds on the scalar coupling from Solar System experiments
\cite{Perivolaropoulos:2009ak,Alsing:2011er}, and let $M_\odot$ be the
mass of the Sun. We get the scaling
\begin{equation}\label{nuunits}
\nu=5.60\times 10^{-21}
\left(\frac{m_s}{m_{20}}\right)^2
\left(\frac{m}{M_\odot}\right)^2\,,
\end{equation}
so that a binary with $m=10^6M_\odot$ (a typical target for eLISA)
could have at most $\nu\sim 5.60\times 10^{-9}$ when $m_s=m_{20}$.
With this definition we can write
\begin{align}
\dot f&=\frac{96}{5}\pi^{8/3}{\cal M}^{5/3}f^{11/3}\left\{1-\frac{2}{3}
\xi(s_1+s_2-2s_1s_2)-\frac{1}{2}\xi\right.\nn\\
&\left.+\left[\frac{\xi\Gamma^2}{12}\left(1
-\frac{\nu}{2(\pi mf)^2}
+\frac{\nu^2}{16 (\pi mf)^4}\right)
\Theta(2\pi f-m_s)
\right.\right.\nn\\
&\left.\left.
+\frac{5 \xi {\cal S}^2}{48}\left(\f{1}{(\pi mf)^{2/3}}-\f{\nu}{(\pi mf)^{8/3}}\right)
\Theta(\pi f-m_s)\right]+\dots\right\}\,.
\end{align} 
Furthermore, we note that (dubbing either $f_0=m_s/(2\pi)$ or $f_0=m_s/\pi$)
\begin{align}
&\int_{-\infty}^fdf'(f')^n\Theta(f'-f_0)=\Theta(f-f_0)\int_{f_0}^fdf'(f')^n\nn\\
&=\Theta(f-f_0)\frac{f^{n+1}}{n+1}+{\rm constant}\,,
\end{align}
where the constant can be absorbed in ``time at coalescence'' and
``phase at coalescence'' integration constants $(t_c,\,\phi_c)$. The
final result for the SPA phase reads
\begin{align}
\psi(f)&=
2\pi ft_c-\phi_c-\frac{\pi}{4}+
\frac{3}{128(\pi {\cal M}f)^{5/3}}\times
\nn\\
&\times\Big\{
1+\zeta
\nn\\
&+\frac{20}{9}A\eta^{-2/5}(\pi{\cal M}f)^{2/3}-16\pi\eta^{-3/5}(\pi{\cal M}f)+\dots
\nn\\
&
+\xi\Gamma^2\nu
\left[\frac{5}{462}\eta^{6/5}(\pi{\cal M}f)^{-2}
-\frac{\nu}{1632}\eta^{12/5}(\pi{\cal M}f)^{-4}\right]
\nn\\
&\times \Theta(2\pi f-m_s)
\nn\\
&+\xi{\cal S}^2\left[\frac{25\nu}{1248}\eta^{8/5}(\pi{\cal M}f)^{-8/3}
-\frac{5}{84}\eta^{2/5}(\pi{\cal M}f)^{-2/3}\right]
\nn\\
&\times \Theta(\pi f-m_s)
\Big\}\,,
\label{spafeta2}
\end{align}
where we defined
\begin{equation}
\zeta=
\frac{2}{3}\xi\left(s_1+s_2-2s_1s_2\right)+\frac{\xi}{2}
-\frac{\xi\Gamma^2}{12} \Theta(2\pi f - m_s)
\,.
\end{equation}

Adding higher-order corrections in the post-Newtonian velocity
parameter $v=(\pi mf)^{1/3}=(\pi {\cal M}f)^{1/3}\eta^{-1/5}$ to the
standard GR phase is trivial.  Eq.~(2.2) of \cite{Berti:2004bd} lists
all contributions up to 2PN (including spin-orbit and spin-spin
interactions). Here we ignore spins, but we do include all nonspinning
contributions up to 3.5PN order: cf. Eq.~(3.3) and (3.4) of
\cite{Arun:2004hn}.

Arun recently argued on general grounds that dipolar emission in
generic extensions of general relativity should always introduce a
term proportional to $v^{-2}$ in the SPA \cite{Arun:2012hf}. Our
calculation shows that a nonzero mass of the scalar introduces
additional structure in the waveform: in particular,
Eq.~(\ref{spafeta2}) contains terms proportional to $\nu$ (and hence
to $m_s^2$) that scale like $v^{-8}$ and $v^{-6}$, and terms
proportional to $\nu^2$ that scale like $v^{-12}$. Arun's argument to
constrain dipolar radiation \cite{Arun:2012hf}, while correct when
$m_s=0$, is not general enough to encompass all scalar-tensor theories
(let alone theories whose action contains quadratic or higher-order
terms in the curvature, such as Gauss-Bonnet or Chern-Simons modified
gravity \cite{Yagi:2011xp,Chatziioannou:2012rf}).  While these large
negative powers of $v$ could in principle be strongly dominant over
``standard'' quadrupolar radiation at small frequencies, the presence
of the Heaviside functions protects the SPA phase from a possible
``infrared divergence'' in the limit $v\to 0$.

Arun pointed out that scalar-tensor theories cannot be easily
incorporated within the original parametrized post-Einsteinian
framework because of additional terms that should appear as amplitude
corrections to the waveform \cite{Arun:2012hf}. This issue was
recently addressed in \cite{Chatziioannou:2012rf}, where the
parametrized post-Einsteinian framework was extended to allow for
amplitude corrections. The problem here is of a different nature,
since we work in the restricted post-Newtonian approximation and
therefore we ignore amplitude corrections ``by construction''. What
happens instead is that some {\em phase} corrections are multiplied by
Heaviside functions $\Theta(\pi f-m_s)$ and $\Theta(2\pi f-m_s)$ that
reflect the dipolar or quadrupolar nature of the various contributions
to the flux. This is a conceptual difficulty with power-law
parametrizations of deviations from general relativity, but it is not
an obstacle in practice, because we know from Solar System
observations that $m_s$ must be small. Therefore in our parameter
estimation calculations we will assume that $m_s=0$, and look for
deviations of $m_s$ from zero. Under this approximation the condition
$f\ge m_s/\pi$ is always satisfied, and all Heaviside functions can be
set equal to one.

In massive scalar-tensor theories, corrections to general relativity
depend on two small parameters, $\nu$ and $\xi$. The structure of the
phasing function (\ref{spafeta2}) has a definite hierarchy when seen
as a multivariate Taylor series in $\xi$ and $\nu$. However, the term
proportional to $v^{-12}$ (which is proportional to $\xi\nu^2$ and
formally of ``third order'' in the small parameters) will produce
sizeable corrections whenever $\nu\sim v^6$.  More generally, it is
possible that some extensions of general relativity will contain
additive corrections to the phasing of the form (say) $\alpha
v^{-a}+\beta v^{-b}$, with $\alpha\ll 1$, $\beta\ll 1$, $a>0$ and
$b>0$.  The parametrized post-Einsteinian framework relies on
identifying a leading-order correction to a general relativistic
waveform. In such cases it may be hard to identify a single
leading-order term: a simple power counting in $v$ is not sufficient,
unless we have independent means of constraining the relative
magnitude of the small parameters $\alpha$ and $\beta$.

\section{Fisher matrix in massive scalar-tensor theories}
\label{sec:Fisher}

Following the notation and terminology of \cite{Berti:2004bd}, we work
in the angle-averaged approximation and we write the gravitational
waveform in the restricted post-Newtonian approximation (where only
the leading-order term is kept in the wave {\em amplitude}) as
\begin{equation}
\tilde h(f)=\gamma {\cal A}f^{-7/6}e^{{\ii}\psi(f)}\,.
\end{equation}
Here ${\cal A}=\frac{1}{\sqrt{30}\pi^{2/3}}\frac{\cal M}{D_L}$, $D_L$
is the luminosity distance to the source and $\gamma$ is a geometrical
correction factor: $\gamma=1$ for Earth-based detectors, and
$\gamma=\sqrt{3}/2$ for Classic LISA and eLISA
(cf.~\cite{Berti:2004bd}). When the binary members are nonspinning the
gravitational waveform depends on seven parameters:
$\{\ln{\cal A},t_c,\phi_c,\ln{\cal M},\ln\eta,\xi,\nu\}$\,.

An estimate of the accuracy in determining these parameters can be
obtained using the Fisher matrix formalism. The calculation of the
Fisher matrix requires explicit expressions for the derivative of the
waveforms with respect to the various parameters. 
We refer the reader to \cite{Poisson:1995ef,Berti:2004bd} for a
detailed overview of the Fisher matrix calculation in the present
context, and to Appendix \ref{app:Fisher} for some subtleties in
estimating the error on $m_s$ given the error on $\nu$.  For theories
involving a massive scalar, Eqs.~(2.13) and (2.14) of
\cite{Berti:2004bd} must be modified as follows.
Eq.~(2.13f) becomes
\begin{align}
\frac{\partial\tilde h}{\partial\ln\cal M}
&=-\ii\frac{5}{128(\pi {\cal M}f)^{5/3}}\tilde h
\nn\\
&\times\Big\{
\xi\Gamma^2\nu
\left[\frac{1}{42}v^{-6}-\frac{\nu}{480}v^{-12}\right]
\Theta(2\pi f-m_s)
\nn\\
&+\xi{\cal S}^2
\left[\frac{5\nu}{96}v^{-8}-\frac{1}{12}v^{-2}\right]
\Theta(\pi f-m_s)
\nn\\
&+1+\zeta
+A_4v^2+B_4v^3+C_4v^4
\Big\}\,,
\end{align}
and Eq.~(2.13g) becomes
\begin{align}
\frac{\partial\tilde h}{\partial\ln\eta}
&=-\ii\frac{1}{96(\pi {\cal M}f)^{5/3}}\tilde h
\nn\\
&\times\Big\{
\xi\Gamma^2\nu
\left[-\frac{9}{308}v^{-6}+\frac{9\nu}{2720}v^{-12}\right]
\Theta(2\pi f-m_s)
\nn\\
&+\xi{\cal S}^2
\left[-\frac{15\nu}{208}v^{-8}+\frac{3}{56}v^{-2}\right]
\Theta(\pi f-m_s)
\nn\\
&+A_5v^2+B_5v^3+C_5v^4
\Big\}\,.
\end{align}

Note that only terms proportional to $v^{-2}$ survive as $m_s\to 0$
(or equivalently as $\nu\to 0$), and they correspond to the $K_4$ and
$K_5$ terms in \cite{Berti:2004bd}.  In the absence of spins our
waveform now depends on seven parameters (rather than six, as in the
massless case), so we have two equations replacing (2.13e) of
\cite{Berti:2004bd}:
\begin{align}
\frac{\partial\tilde h}{\partial\xi}
&=\ii\frac{3}{128(\pi {\cal M}f)^{5/3}}\tilde h
\nn\\
&\times\Big\{
\Gamma^2
\left[\frac{5\nu}{462}v^{-6}-\frac{\nu^2}{1632}v^{-12}-\f{1}{12}\right]
\Theta(2\pi f-m_s)
\nn\\
&+{\cal S}^2
\left[\frac{25\nu}{1248}v^{-8}-\frac{5}{84}v^{-2}\right]
\Theta(\pi f-m_s)\nn\\
&+\frac{2}{3}\left(s_1+s_2-2s_1s_2\right)+\frac{1}{2}
\Big\}\,,\label{hxi}\\
\frac{\partial\tilde h}{\partial\nu}
&=\ii\frac{3}{128(\pi {\cal M}f)^{5/3}}\tilde h
\nn\\
&\times\Big\{
\xi\Gamma^2
\left[\frac{5}{462}v^{-6}-\frac{\nu}{816}v^{-12}\right]
\Theta(2\pi f-m_s)
\nn\\
&+\xi{\cal S}^2
\left[\frac{25}{1248}v^{-8}\right]\Theta(\pi f-m_s)
\Big\}\,.\label{hnu}
\end{align}

We remark again that we are interested in deviations of $m_s$ and
$\xi$ from zero. Following the procedure outlined after Eq.~(2.15) in
\cite{Berti:2004bd}, when computing the Fisher matrix and its inverse
we set $m_s=\nu=0$ in the derivatives listed above (of course, all
Heaviside functions are then equal to one). We verified the validity
of this assumption by repeating the calculation with $m_s=10^{-20}$~eV
(a mass for which we know from Solar System constraints that the
coupling must be small \cite{Alsing:2011er}): our bounds are
essentially unaffected by this modification.  However it should be
remarked that it is {\it not} possible to set $\xi=0$. The reason is
apparent from Eq.~(\ref{hnu}): $\frac{\partial\tilde h}{\partial\nu}$
is proportional to $\xi$, and therefore the Fisher matrix becomes
singular in the limit $\xi\to 0$ ($\omega_{\rm BD}\to \infty$).

By inspection we see that when $\nu$ contributes to the SPA phase in
Eq.~(\ref{spafeta2}) it is always multiplied by the coupling parameter
$\xi$, so we can only constrain the product $\xi\nu$, or in other
words the combination $\delta \nu/\omega_{\rm BD}$, where $\delta \nu$
measures variations of $\nu$ from zero. In practice we can estimate
the accuracy in estimating $\delta\nu$ by setting $\xi=(2+\omega_{\rm
  BD})^{-1}$ equal to some small but nonzero value compatible with
Solar System bounds. For the calculations in this paper we chose
$\omega_{\rm BD}=4\times 10^5$, but we verified that the measurement
error on $\delta \nu$ scales linearly with $1/\xi$ for small coupling
parameters. Therefore we present our results in terms of the
combination $\delta \nu/\omega_{\rm BD}$, which is independent of the
assumed value of $\omega_{\rm BD}$ in the small-coupling limit.

Appendix A shows that once we know the statistical error on $\nu$, say
$\sigma_\nu$, the statistical error $\sigma_{m_s}$ on $m_s$ can be
estimated as
\be 
\sigma_{m_s}\simeq \f{\sigma_\nu^{1/2}}{m}\,.
\label{sigms}
\ee

\begin{figure}[thb]
\includegraphics[width=8.5cm,clip=true]{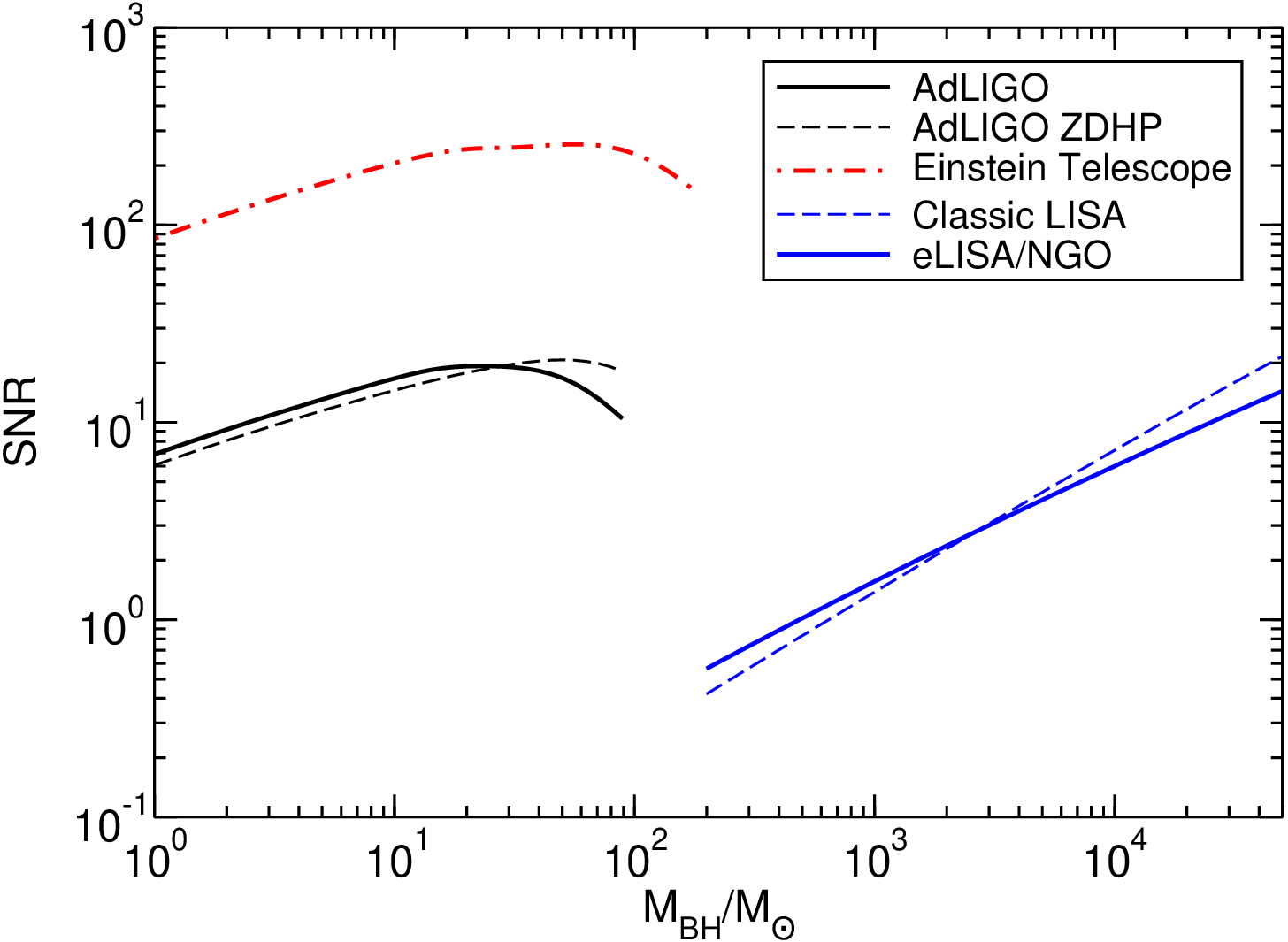}
\caption{SNR for observations of neutron star-black hole binaries as a
  function of the black hole mass $M_{\rm BH}$ for various detectors
  (AdLIGO, AdLIGO ZDHP, ET, Classic LISA and eLISA/NGO) at luminosity
  distance $D_L=200$~Mpc.}
\label{fig:SNR}
\end{figure}
\begin{figure*}[htb]
\includegraphics[width=8.5cm,clip=true]{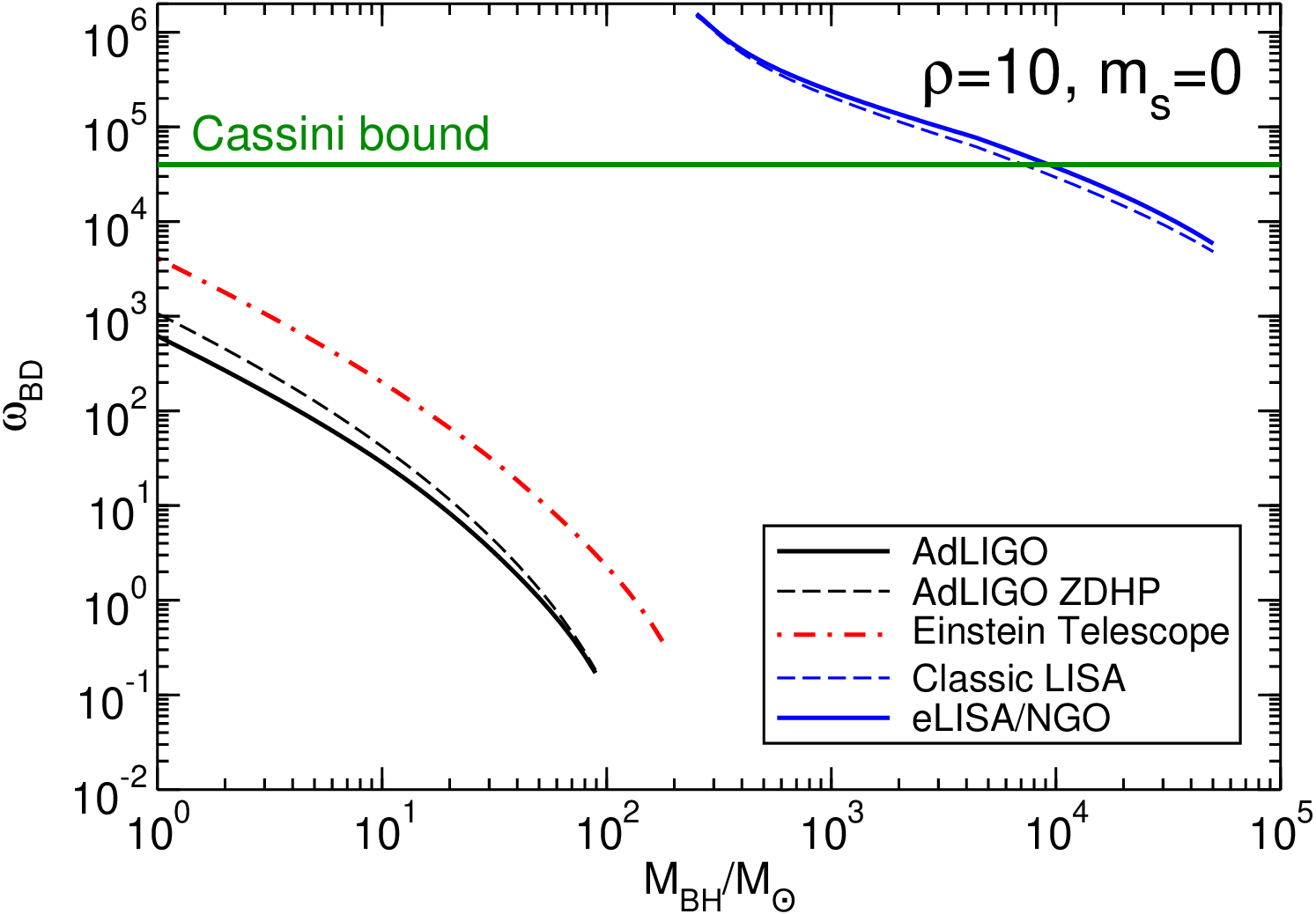}
\includegraphics[width=8.5cm,clip=true]{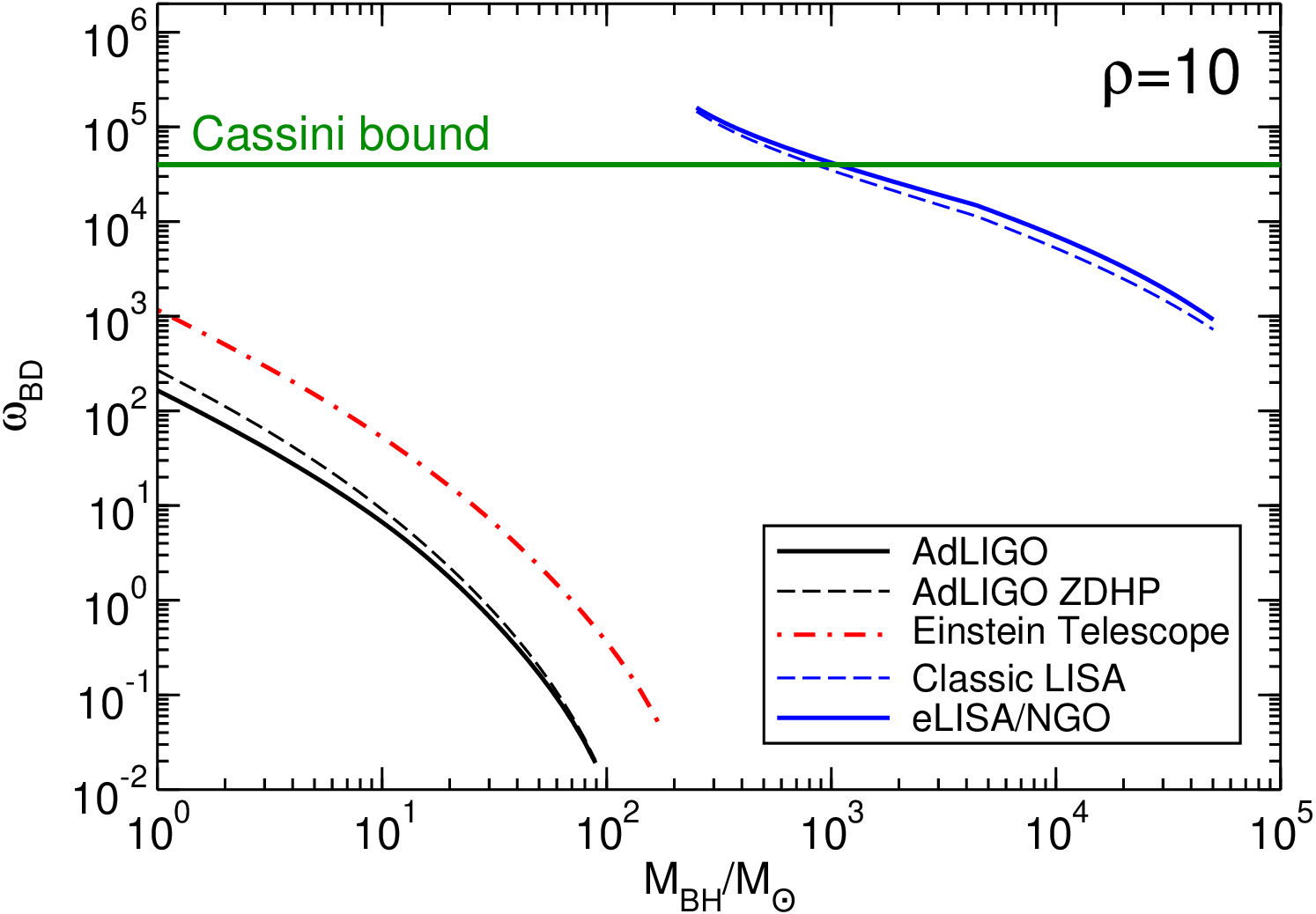}
\caption{Bound on $\omega_{\rm BD}$ obtained and considering a
  six-parameter Fisher matrix at fixed SNR $\rho=10$.  Left: we set
  $m_s\propto \nu=0$ and consider a six-parameter Fisher
  matrix. Right: same, but for a seven-parameter Fisher matrix with
  $\nu\neq 0$.}
\label{fig:bounds}
\end{figure*}

\section{Bounds with Earth- and space-based interferometers}
\label{sec:bounds}

For each of the five detectors (AdLIGO, AdLIGO ZDHP, the Einstein
Telescope, Classic LISA and eLISA) we present ``conservative''
estimates of the bounds by considering binaries with fixed SNR
$\rho=10$. There are two reasons for this choice: one is conceptual
(Fisher matrix calculations are unreliable for black hole masses such
that $\rho \lesssim 10$) and one is practical (using $\rho=10$ allows
direct comparison with previous work, e.g. \cite{Berti:2004bd}).

We consider neutron star-black hole binaries where the neutron star
has mass $M_{\rm NS}=1.4M_\odot$, and we vary the black hole mass in a
range depending on the optimal sensitivity window of each detector.
Bounds are inversely proportional to the SNR, which in turn is
inversely proportional to the luminosity distance of the binary
($\rho\sim 1/D_L$). To facilitate rescaling of our results, in
Fig.~\ref{fig:SNR} we plot the SNR of neutron star-black hole binaries
at luminosity distance $D_L=200$~Mpc as a function of the black hole
mass $M_{\rm BH}$, for all five detectors\footnote{The neutron
  star-black hole binaries of interest here have large mass ratio. For
  large mass-ratio systems the SNR is proportional to $M_{\rm
    NS}/M_{\rm BH}$. Bounds on the scalar mass and on $\omega_{\rm
    BD}$ scale linearly with the SNR, so changing the mass of the
  neutron star has the trivial effect of rescaling the bounds by a
  constant factor which is very close to unity.}. Note that AdLIGO and
AdLIGO ZDHP are very similar in terms of SNR. The same applies to
Classic LISA and eLISA: in fact, eLISA has slightly larger SNR when
$M_{\rm BH}\lesssim 2000~M_\odot$, mainly because the eLISA
``armlength'' is a factor of five smaller with respect to Classic LISA
(see \cite{AmaroSeoane:2012km}).

The bounds that can be placed on $m_s/\sqrt{\omega_{\rm BD}}$ were
discussed in Section \ref{sec:intro}. In Fig.~\ref{fig:bounds} we
complement those results by plotting the bounds on $\omega_{\rm BD}$,
estimated as in \cite{Berti:2004bd}. The left panel shows the bounds
on $\omega_{\rm BD}$ that we would obtain if we considered {\em
  massless} scalar tensor theories, as in \cite{Berti:2004bd}. In the
right panel we show that when $m_s\neq 0$ the bounds get worse by
about one order of magnitude. This is expected: we are adding one
additional, highly correlated parameter to the waveform, and this
reduces parameter estimation accuracy on {\em all} intrinsic
parameters of the binary by roughly one order of magnitude. This
degradation of the bounds is analogous to what happens when we add
spin-orbit terms to the SPA: cf. the discussion around Table IV of
\cite{Berti:2004bd}. As we could expect from the SNR plots, AdLIGO
bounds are very similar to AdLIGO ZDHP bounds, and eLISA does slightly
better than LISA for small black hole masses. Whether we consider
massive or massless scalar-tensor theories, the best bounds
(competitive with the Cassini bound) would come from observations of
the intermediate mass-ratio inspiral of a neutron star into a black
hole of mass $M_{\rm BH}\lesssim 10^3~M_\odot$, as observed by a
space-based instrument such as eLISA or Classic LISA.

\section{Conclusions and outlook}
\label{sec:conclusions}

In this paper we have studied the bounds on massive scalar tensor
theories with constant coupling from gravitational-wave observations
of quasicircular, nonspinning neutron star-black hole binary inspirals
in the restricted post-Newtonian approximation. We found that neutron
star-black hole systems will yield bounds $(m_s/\sqrt{\omega_{\rm
    BD}})(\rho/10)\lesssim 10^{-15}$, $10^{-16}$ and $10^{-19}$~eV for
Advanced LIGO, ET and Classic LISA/eLISA, respectively. We also found
that the best bounds on $\omega_{\rm BD}$ would come from space-based
observations of the intermediate mass-ratio inspiral of a neutron star
into a black hole of mass $M_{\rm BH}\lesssim 10^3~M_\odot$.

It would be interesting to drop the restricted post-Newtonian
approximation and to consider amplitude corrections in the context of
massive scalar-tensor theories
(cf.~\cite{Arun:2012hf,Chatziioannou:2012rf} for recent work in this
direction). Furthermore, in our analysis we have neglected time delay
effects which arise because the massive scalar modes propagate slower
than the massless tensor modes; while presumably small, it is
worthwhile to investigate how these effects would change our bounds.
Another obvious extension of this work would be to include spin
precession, orbital eccentricity and the merger/ringdown waveform. All
of these effects could improve our conservative estimate of the bounds
on $m_s$ and $\omega_{\rm BD}$. Finally, it would be interesting to
estimate the improvement on the bounds that would result from
observing {\it several} neutron star-black hole systems with one or
more detectors (see e.g.  \cite{Gair:2010yu,Berti:2011jz} for
preliminary studies in a slightly different context).

\noindent
{\bf \em Acknowledgments.}
E.B. was supported by NSF Grant No. PHY-0900735 and NSF CAREER Grant
No. PHY-1055103. We also acknowledge support from the NRHEP--295189
FP7--PEOPLE--2011--IRSES Grant and by FCT projects
PTDC/FIS/098032/2008 and PTDC/FIS/098025/2008. We are grateful to
K.~G.~Arun, Giulio D'Agostini, Paolo Pani, Nicol\`as Yunes and
Clifford Will for discussions. We are particularly grateful to Michele
Vallisneri for clarifying some issues related to error estimates.

\appendix

\section{Some subtleties in the calculation of parameter estimation errors}
\label{app:Fisher}

The inversion of the Fisher matrix yields the statistical errors on
the seven parameters chosen to parametrize the waveform, namely
$\{\ln{\cal A},t_c,\phi_c,\ln{\cal M},\ln \eta,\xi,\nu\}$, and the
associated correlation coefficients. Our goal is to get a bound on
$m_s$, assuming that the background value of $m_s$ is zero. 

In order to evaluate the statistical error of the scalar field mass
$m_s$, we first find the statistical error of the variable
\begin{equation}
m_s^2=\frac{\nu}{m^2}=\frac{\nu \eta^{6/5}}{{\cal M}^2}
\end{equation}
as a function of the statistical errors of $\nu,{\cal M},\eta$.

The error on a parameter $\theta^a$ can be obtained by inverting the
Fisher matrix
\be
\Gamma_{ab}=\left(\f{\p h}{\p \theta^a}\Big|\f{\p h}{\p \theta^b}\right)\,.
\ee
If we consider the logarithm of a parameter, say $\theta^{\bar a}=\ln
\theta^a$, as we do here for ${\cal M}$ and $\eta$, then obviously
\be
\Gamma_{\bar ab}=
\left(\f{\p h}{\p \ln \theta^a}\Big|\f{\p h}{\p \theta^b}\right)=
\theta^a \Gamma_{ab}\,.
\ee
So switching to the logarithm of a parameter corresponds to a
rescaling of the corresponding matrix element by $\theta_a$. For the
elements of the variance-covariance matrix $\Sigma^{ab}$ we have
\be
\Sigma^{\bar a \bar a}=\Gamma_{\bar a \bar a}^{-1}=(\theta^a)^{-2} \Gamma_{aa}^{-1}\,,
\quad
\Sigma^{\bar a b}=(\theta^a)^{-1} \Gamma_{ab}^{-1}\,.
\ee
The transformation of the variance-covariance matrix under a change of
variables is given (e.g.) in Appendix A of
\cite{Sesana:2010wy}. Suppose we are given functional relations
$y_i(x_j)$, and introduce the Jacobian $D_x^{ij}=\f{\p y_i}{\p
  x_j}$. Following the notation of that paper, the result reads
\be
\Sigma_x=
(D_x)^{-1}
\Sigma_y
((D_x)^T)^{-1}\,.
\ee
For reasons explained in Appendix A of \cite{Sesana:2010wy}, by
considering the diagonal components of the error matrix in the new
variables $y$ we get a conservative estimate of the errors. In
conclusion, if $x_i=(\nu\,,\ln {\cal M}\,, \ln\eta)$ we can write
\begin{align}
\sigma_{m_s^2}^2 &= 
\Sigma_{m_s^2\, m_s^2}=\sum_{ij} 
\f{\p m_s^2}{\p x_i}
\f{\p m_s^2}{\p x_j}
\Sigma_{ij}= \\
&=
\f{\eta^{12/5}}{{\cal M}^4}\left[
\Sigma_{\nu\nu}+
\f{36}{25}\nu^2 \Sigma_{\ln \eta\,\ln \eta}+
4\nu^2 \Sigma_{\ln {\cal M}\,\ln {\cal M}}\right.\nn\\
&+\left.
\f{12}{5}\nu \Sigma_{\nu\,,\ln \eta}-
4\nu \Sigma_{\nu\, \ln{\cal M}}-
\f{24}{5}\nu^2 \Sigma_{\ln \eta\,\ln {\cal M}}
\right]\,.\nn
\end{align}
Because we assume that $\nu=0$ in the background, this expression
reduces to
\begin{align}
\sigma_{m_s^2} 
= \f{\sqrt{\Sigma_{\nu\nu}}}{m^2}
= \f{\sigma_{\nu}}{m^2}\,.
\label{propagation}
\end{align}
Following common use in the gravitational-wave literature, we will
ignore subtle issues associated with the fact that $m_s>0$ and simply
use $\sigma_{m_s}=\sqrt{\sigma_{m_s^2}}$ to obtain the conservative
estimate given in Eq.~(\ref{sigms}) of the main text: if there is a
$68\%$ probability of finding $m_s^2<\sigma_{m_s^2}$, then there is
the same probability of finding $m_s<\sqrt{\sigma_{m_s^2}}$.

%

\end{document}